\documentclass{article}
\usepackage{amssymb,amsmath}
\usepackage{times}
\usepackage{graphicx}
\usepackage{amssymb}

\begin{document}

\centerline{\Large{Attosecond quantum entanglement in neutron
Compton scattering}} \vspace{0.3cm}
 \centerline{\Large{from water in the keV range}}


\vspace{1cm} \centerline{C. A. Chatzidimitriou-Dreismann$^1$}

\vspace{1cm} \noindent $^1$Institute of Chemistry, Faculty II,
Technical University of Berlin, D-10623 Berlin, Germany. \ \
\textit{Email:} dreismann@chem.tu-berlin.de

\vspace{1cm} \noindent \abstract{Scattering of neutrons in the
24--150 keV incident energy range from H$_2$O relative to that of
D$_2$O and H$_2$O--D$_2$O mixtures was reported very recently.
Studying time-of-flight integrated intensities, the applied
experimental procedure appears to be transparent and may open up a
novel class of neutron experiments regarding the "anomalous"
scattering from protons, firstly observed  in our experiment at
ISIS  in the 5--100 eV range. The keV-results were analyzed within
standard theory, also including (1) multiple scattering and (2)
the strong incident-energy dependence of the neutron-proton cross
section $\sigma_H(E_0)$ in this energy range. The analysis reveals
a striking anomalous ratio of scattering intensity of H$_2$O
relative to that of D$_2$O of about 20\%, thus being in
surprisingly good agreement with the earlier results of the
original experiment at ISIS.}\\


\section{Introduction}
Several neutron Compton scattering (NCS) experiments on liquid and
solid samples containing protons or deuterons show a striking
anomaly, i.e.  a shortfall in the intensity  of energetic neutrons
scattered by the protons; cf.~\cite{PRL97,NCS,PRL2003,highlights}.
E.g., neutrons colliding with water for just $100-500$
attoseconds (1\,as = $10^{-18}$\,s) will see a ratio of hydrogen
to oxygen of roughly 1.5 to 1, instead of 2 to 1 corresponding to
the chemical formula H$_2$O.
 Due to the large energy and momentum transfers
applied, the duration of a neutron-proton scattering event (the
so-called scattering time) is a fraction of a femtosecond  which
is extremely short compared to the usual condensed-matter
relaxation times.

 Recently this new effect has been independently confirmed
 by electron-proton Compton scattering (ECS) from a solid
polymer \cite{PRL2003,highlights,NIMB}. The similarity of ECS and
NCS results is striking because the two projectiles interact with
protons via fundamentally different forces, i.e.~the
electromagnetic and strong forces.

Due to its novelty and far-reaching consequences, however, this
effect has been the focus of various criticisms,
cf.~\cite{Blostein,Cowley}. Therefore,
 considerable work to identify possible sources of experi\-men\-tal
and data analysis errors was made during the last five years,
which succeeded to demonstrate the excellent working conditions of
the spectrometer Vesuvio; see \cite{J+T} for an account in detail.
Extending these investigations,  the complete "exact formalism" of
data analysis \cite{Blostein} was applied to NCS-data by Senesi et
al. \cite{Senesi}, for the first time.
 Analysis of time-of-flight (TOF) spectra from solid HCl
 revealed the existence of an "anomalous" decrease
of the scattering intensity from H by 34\%. Moreover, this result
was found to be in excellent agreement with the corresponding
outcome of  the standard data analysis procedure applied at ISIS
\cite{Senesi}.
 Additionally, the mentioned standard NCS-data analysis method
\cite{J+T} was successfully compared with a newly proposed (by B.
Dorner, ILL) model-free data-analysis procedure, the latter being
independent of the form of the momentum distribution and the
resolution function \cite{PRBDorner}.

\section{NCS in the keV range}
In a novel experiment by Moreh et al.~\cite{Moreh}, scattering
results of neutrons  from H$_2$O relative to that of D$_2$O and
H$_2$O--D$_2$O mixtures were reported, in the incident energy
range  about 24--150 keV.
 This
experiment was carried out to search for the aforementioned
anomalous decrease \cite{PRL97} in the neutron scattering
intensity from protons (relative to that from deuterons) at
interaction times in the attosecond regime. It is important to
note that the energy range here is about 3000 times larger than
that of our original neutron Compton scattering (NCS) experiment
carried out with the electron volt spectrometer eVS (newly:
Vesuvio) at the neutron spallation source ISIS \cite{PRL97}.

In clear contrast to the original NCS results \cite{PRL97}, Moreh
et al.~claimed that the results in keV range exhibited no
anomalous behavior.  It was concluded
 that within an overall statistical accuracy of 3\% there is no
evidence for any deviation from the ratios of scattering
intensities conventionally calculated on the basis of the
tabulated total neutron cross sections \cite{Moreh}. In
addition,  it was correctly argued that one would have to shake
some well established notions in physics to explain the
aforementioned scattering anomaly.

Our theoretical analysis of these experimental data, and in
particular their comparison with predictions of standard NCS
theory \cite{Watson}, is considered in the following. This
analysis reveals a striking anomalous ratio of scattering
intensity of H$_2$O relative to that of D$_2$O of about 20\%.
 Extending the first analysis of single scattering events \cite{antiM},
it will be shown that
 $(a)$ neither multiple scattering
 $(b)$ nor the strong dependence of the proton total cross section $\sigma_H(E_0)$
on incident energy $E_0$, in the considered keV range,  do
considerably affect the anomaly under investigation.

\subsection{Experimental}
 First of all, it is important to notice that both setups, i.e. that
of the new keV-neutron experiment at the Rensselaer Polytechnic
Institute (RPI) \cite{Moreh}  and that of Vesuvio  at ISIS
\cite{J+T},  are basically similar and thus the interpretation of
their results ought to be based on the same basic theory. The
following related remarks should be helpful:

 $(i)$ Both are so-called \textit{inverse geometry}
 time-of-flight (TOF)
 setups; i.e., the final energy $E_f$ of the measured
neutrons is fixed, the neutron initial energy $E_0$ is
"continuous", and the scattered neutrons are analyzed using a
"filter" (of Fe at RPI, with $E_f=24.3 (\pm 1.1)$ keV) or
"analyzer foil" (of Au at ISIS, with $E_f=4.91 (\pm 0.14)$ eV).

 $(ii)$ The  range of
scattering angles $\theta$ is similar in both setups, (i.e. one
detector integrating over  $ \theta = 25^\circ-65^\circ $ at RPI;
32 detectors measuring at various scattering angles in the range
$35^\circ-67^\circ$ at ISIS).

 $(iii)$ As a consequence, and according to standard theory \cite{Watson},
the characteristic neutron-proton scattering time in the keV range
is shorter by a factor of about 70 \cite{Moreh} as compared to
that of the Vesuvio-setup. In view of some of the mentioned
theoretical models (see below),
however, this difference of scattering times may not preclude the
appearance of the considered effect in the keV range.

 $(iv)$ The Impulse Approximation
(IA) can  be safely assumed to be exact in the keV range
\cite{Moreh}, and it is already known to be sufficiently fulfilled
in the eV range of Vesuvio \cite{J+T,Watson}. In simple terms,
each neutron scatters from a single nucleus (of H, D or O).

$(v)$ The setup at RPI cannot resolve the neutrons scattered
 from different nuclei and yields instead an \textit{integrated}
scattering signal arising from the (O, H and/or D) nuclei of the
liquid samples. The ISIS setup, however, provides a H-recoil peak
well resolved from that of D or O, but this does not represent
any significant difference that would prevent comparison of
results. (E.g., one can simply integrate over the individual peak
intensities, when required).

 $(vi)$ Furthermore, note that the scattering intensity, measured
by a single detector, is represented by the area of the 24.3 keV
line in the TOF spectra, i.e. by adding up the total number of
counts in the 24.3 keV peak \cite{Moreh}. An independent fission
detector, placed inside a separate beam tube, was employed as a
neutron flux monitor and served to normalize the TOF spectra.
After subtracting the background from each signal, the intensity
ratios are taken, as presented in Fig.~3 of \cite{Moreh}. The
measurements were repeated about five times, always giving about
the same results within statistics (personal communication).

$(vii)$ Already here it should be emphasized that, as already
stressed in Ref.~\cite{Moreh}, multiple scattering effects do not
affect the \textit{ratio} of scattering intensities from the two
samples; see also below.

\subsection{Single scattering events}
   Recognizing the crucial importance of these novel keV-experimental
results, we analyzed in  detail \cite{antiM} the data processing
scheme indicated in Ref.~\cite{Moreh} on the basis of standard
scattering theory at large energy transfers \cite{J+T,Watson},
where the IA is valid; cf.~point $(iv)$ above. Some omissions in
\cite{Moreh} have been revealed and then amended, thus leading to
a radical revision of the main finding and conclusion of
Ref.~\cite{Moreh}. In accordance with the preceding point $(vii)$,
in that calculation \cite{antiM} we considered \textit{single}
scattering events only. That calculation \cite{antiM}, which also
took into account the strong dependence of the proton total cross
section $\sigma_H(E_0)$ on incident energy $E_0$
\cite{sigmavalues},  contains no fitting parameter.

A full account in detail of these investigations is presented in
Ref.~\cite{antiM}. The results are summarized in Fig.~1 and lead
to a radical revision of the main finding and conclusion of
Ref.~\cite{Moreh}:
 The correct keV-data reduction reveals a
strongly anomalous ratio of scattering intensity of H$_2$O with
that of D$_2$O  of about 20\%, thus being in surprisingly good
agreement with the associated results of the original
ISIS-experiment \cite{PRL97}.

Our results \cite{antiM} were  confirmed independently by Monte
Carlo calculations of Mayers \cite{Jerry}, applying the routines
for NCS-data analysis available at ISIS. Also these calculations
contain no fitting parameters.

 \begin{figure}[htb]
     \centering
     \includegraphics[width=7cm,height=7cm]{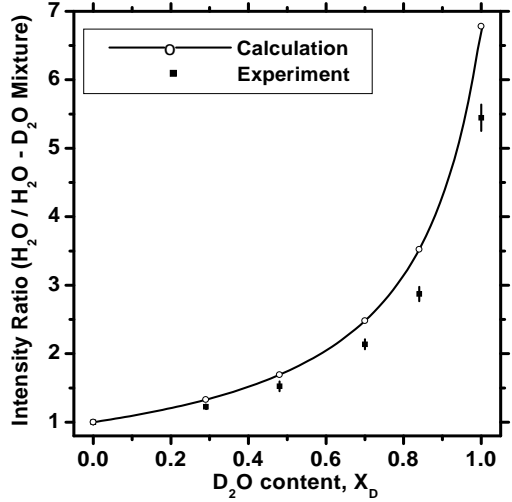}
     \caption{
     Intensities of single scattering events \cite{antiM}.
     Measured (full squares with error bars) --- taken from
Ref.~\cite{Moreh} --- and corrected calculation (solid line, open
circles) for $\theta = 45^\circ$ of scattered intensity ratios
versus $X_D$, the D$_2$O concentration in the H$_2$O-D$_2$O
mixture.
 The discrepancy between experimental data and prediction
of conventional theory is clearly discernible for data with $X_D
\geq 0.5$. The calculated intensity ratio for pure H$_2$O and pure
D$_2$O is ca.~23\% higher than the experimental one.   }
 \end{figure}

\subsection{Double scattering events -- Energy dependence of
$\sigma_H$}

The thickness of all samples in the keV experiment was $d=0.18$
cm. However, as already mentioned under point $(vii)$ above, is
was shown that multiple scattering effects did not affect the
considered ratio of scattering intensities; explicitly:
 "The effect on the \textit{ratio} of scattering intensities from
the two samples is $<$1\% and was neglected.";
  see page 4 of Ref.~\cite{Moreh}.

Applying the Monte Carlo routines available at ISIS, J. Mayers
analyzed the multiple scattering effects on water and D$_2$O, for
the  sample thickness of $d=0.18$ cm \cite{Jerry}. In these
calculations, however, the proton total cross section $\sigma_H$
was kept constant (e.g. 15.7 barn). The obtained results  clearly
showed that,  in the H$_2$O case, the single scattering events are
attenuated due to the considerable sample thickness. (Note that
the transmittance of the H$_2$O sample (for $E_0=48.6$ keV) is ca.
81\%, see numerical calculations below). But, the crucial point
here was that the \textit{twice} scattered neutrons were found
 to go mainly at forward scattering angles,
 and that $\theta \approx 45^\circ$ is at the center of the
 the double scattering events in the forward angle range.
Thus, as sample attenuation  increases, multiple scattering also
increases. As a result, and including multiply scattered neutrons,
the ratio of scattering intensity of H$_2$O relative to that of
D$_2$O was calculated to be about 6.4 \cite{Jerry}, i.e. thus
about 17\% larger than the associated experimental value of $5.5$
\cite{Moreh}.

Nevertheless, and due to the potentially far-reaching consequences
of the provided new keV experimental results, we extended the
aforementioned analysis \cite{antiM} and  Monte Carlo simulations
\cite{Jerry}  by \textit{including the strong dependence of the
proton total cross section $\sigma_H$ on incident energy $E_0$}
\cite{sigmavalues} into the calculations of double scattering.

This dependence could be important, for the following reason.
Neutrons which are not scattered once into the detector are
multiply scattered. Such neutrons may have very high incident
energies $E_0$,  before they end up with final energy $E_f=24.3$
keV  and become measured. Therefore they would have a much lower
cross-section and hence the probability of scattering  for these
neutrons may be very low. Consequently it is conceivable that the
multiple scattering contributions from protons could be
significantly lower than those conventionally calculated on the
basis of a fixed proton-neutron cross section (say, e.g., with
$\sigma_H = 15.7$ barn, being valid for $E_0 = 48.6$ keV).

Here it should  be emphasized that the existing Monte Carlo
routines for the calculation of multiple scattering effects cannot
handle  the present case of  $\sigma_H$  depending on incident
energy $E_0$; cf.~\cite{JerryMultScatt}. The reason for this
should be  that, up to ca.~1 keV, the "free" total cross section
of H remains constant, and the thus far existing NCS
investigations were well within this energy range.

To simplify the derivations and pinpoint their main physical
aspects, we consider here double scattering from the water's
protons only.  Let $N_H$ be the number of H atoms per unit volume
and
 \begin{equation}
\mu(E_x) \equiv N_H \sigma_H(E_x)
 \end{equation}
at "incident" neutron energy $E_x$ before the $x$-th
neutron-proton collision. Let $\beta$ be the scattering angle of
the impinging neutron with initial energy $E_0$ due to the
\textit{first} neutron-proton collision, and $\phi$ being the
associated azimuthal angle. (As usual, polar coordinates are used
here.) Furthermore, let $\chi$ be the angle between the two
neutron-velocity vectors before and after the \textit{second}
scattering event. We consider neutrons with a total scattering
angle $\theta$ and with final energy $E_f=24.3$ keV going into the
detector. As is well known \cite{Vineyard,Agraval}, the
"intermediate" scattering angle $\chi$ depends on both $\beta$ and
$\theta$. A short calculation yields
 \begin{equation}
\cos\chi= \cos\theta\cos\beta+\sin\theta\sin\beta\cos\phi
 \end{equation}
As conventionally, we consider here an infinite slab geometry for
the sample, with thickness $d$, and the incident neutron beam
 in the direction $z$ being perpendi\-cu\-lar to the slab.

According to standard theory \cite{J+T,Lovesey}, the neutron
energy $E_1$ between first and second neutron-proton collision is
\begin{equation}
 E_1 = \frac{E_f}{\cos^2\chi}
 \end{equation}
and the associated initial energy is
\begin{equation}
 E_0 = \frac{E_f}{\cos^2\chi\,\cos^2\beta}
\label{E0}
 \end{equation}
Recall that here $E_f=24.3$ keV. Thus, for some double-scattering
events, $E_0$ may be very large and  the cross section
$\sigma_H(E_0)$ for the first scattering process  very small
\cite{sigmavalues}.

Furthermore we need the three attenuation factors (also called
self shielding or absorption corrections) \cite{Windsor} for  the
neutron propagation through the sample:
\begin{equation}
U_0  = \exp[-\mu(E_0)\,z_0]
 \end{equation}
$z_0$ being the neutron path in the sample until its first
collision,
\begin{equation}
          U_{1} = \exp[-\mu(E_1)\,s_1]
 \end{equation}
$s_1$ being the neutron path in the sample between its first and
second collision, and finally
\begin{equation}
 U_f  = \exp\left[-\mu(E_f)\,\frac{d-z_0 -
          s_1cos\beta}{\cos\theta} \right]
 \end{equation}
for the path between the second collision and the exit from the
sample, in the fixed scattering direction $\theta$ towards the
detector.

Furthermore, the probability for neutron scattering under
scattering angle $\psi$ in the neutron-proton collision is
proportional to $\cos\psi$; see Eq.~(1.88) of \cite{Lovesey}.

The intensity of impinging neutrons in the keV experiment is
proportional to $I(E_0)= E_0^{-0.65}$ \cite{Moreh}.

As a result, the above partial processes lead straightforward to
the following expression for the probability $P_{dse}$ of a
double-scattered neutron  with final energy $E_f=24.3$ keV
measured with the detector at scattering angle $\theta$:
$$
P_{dse}  \propto \int I(E_0)\,[U_0\,\mu(E_0)\sin\beta\cos\beta] \
\ \ \ \ \ \ \ \ \ \ \ \ \ \ \ \
$$
\begin{equation}
\hspace{1.5cm}\times [U_{1}\,\mu(E_1)\,\cos\chi]\,U_f\,
 dz_0 \,d\beta\, d\phi \,ds_1
  \label{dse}
 \end{equation}
 (The squared brackets are for convenience only; they include the
factors belonging to each scattering event.)
 The factor $\sin\beta$ is due to the  scattering
 after the first collision \cite{Lovesey}.
Note the presence of the factors $\mu(E_0)$ and $\mu(E_1)$, which
are associated with the scattering probability at the two
scattering points.

 \begin{figure}[htb]
     \centering
     \includegraphics[width=7.5cm,height=6.5cm]{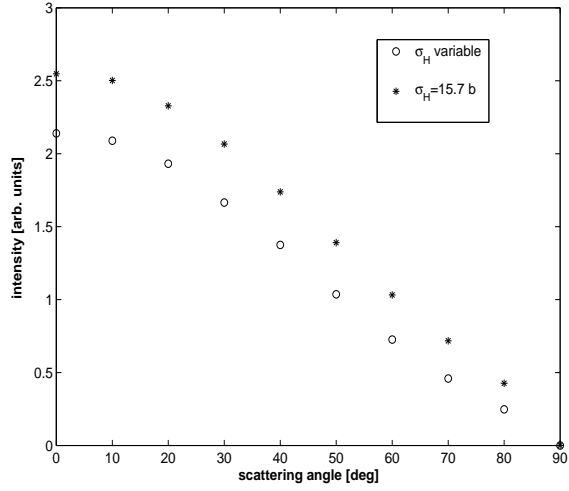}
     \caption{
Intensity of double scattered neutrons from protons of pure H$_2$O
as a function of total scattering angle $\theta$, calculated with
Eq.~(\ref{dse}).     Asterisks: Calculation with fixed cross
section $\sigma_H =$15.7 b.
 Circles: Energy-dependent $\sigma_H(E_0)$, with $E_0$ determined
 by Eq.~(\ref{E0}). Note the "forward" orientation of the twice
scattered neutrons, which is equal in both calculations. The
effect of the energy dependence of $\sigma_H(E_0)$ leads to a
decrease of the twice scattered neutrons of about 23 \% at
$\theta=45^\circ$. This amounts to ca. 3.5\% of the complete
intensity only; see the text.}
 \end{figure}

\subsection{Main result -- strongly anomalous $R$}

Note that all these equations are valid for the case of
energy-dependent $\sigma_H(E_0)$. The probabilities for the twice
scattered neutrons can be calculated with Eq.~(\ref{dse}). The
evaluation can be carried out with the Monte Carlo integration
method; see e.g. Sect.~7.6 of \cite{NumRec}.

 The following
observation is now of crucial importance. The same equation
(\ref{dse}) can also be evaluated for the case of \textit{fixed
cross section} $\sigma_H =$const. (For this, one simply has to fix
the value of $\sigma_H$ in Eq.~1.)
  Thus the results of both
calculations can be directly compared, as they are shown in
Fig.~2. This comparison shows \textit{quantitatively} the effect
of the energy dependence of $\sigma_H$, for various scattering
angles. The results obtained with fixed $\sigma_H$-value confirm
the aforementioned Monte Carlo calculations of Mayers. The effect
of the energy dependence $\sigma_H(E_0)$ leads, at
$\theta=45^\circ$,  to a slight decrease of the twice scattered
neutrons of about 23 \%  only.

Furthermore, calculation of the single-scattered neutron intensity
at $\theta$ was carried out by
\begin{equation}
  P_{sse}  \propto \int_{0}^{d} I(E_0)\,[U'_0\,\mu(E'_0)\cos\theta] \,U'_f\,
   dz_0
  \label{sse}
 \end{equation}
with $E'_0 = E_f/ \cos^2 \theta$, $U'_0  = \exp[-\mu(E'_0)\,z_0]$
and
\begin{equation}
U'_f  = \exp\left[-\mu(E_f)\,\frac{d-z_0}{\cos\theta} \right]
 \end{equation}
(This integral can be calculated analytically, too
\cite{Windsor}.) Comparison of the above results showed that the
ratio of double- to single-scattered neutrons is about 0.15 at
$\theta=45^\circ$. Thus we may proceed to the following crucial
conclusion: If one approximately neglects higher-order scattering
events (which were estimated to be about 2-3\% of the total
scattering), then the aforementioned 23\%-reduction of the
double-scattered neutrons amounts to ca. 3.5\% of the complete
intensity at $\theta=45^\circ$. This furthermore implies that --
neglecting double-scattering from D$_2$O  -- the earlier
calculated ratio $R=6.78$ \cite{antiM}    should be slightly
reduced to $R=6.55$, which is still about 19\%, and thus
significantly, larger than the experimentally measured ratio
$R_{exp}=5.5$

Moreover, if one would also take into account double scattering
events from D$_2$O, it is obvious that the considered ratio $R$ of
integrated scattering intensities from H$_2$O and D$_2$O will be
slightly increased again. Thus we may safely conclude that the
theoretically predicted value of the ratio $R$ still remains about
20\% larger than the experimental one $R_{exp}=5.5$.

This crucial result is fully in line with the previous
calculations based on single scattering events only \cite{antiM}.
It shows that inclusion of double scattering, as well as the
strong dependence of the proton total cross section $\sigma_H$ on
incident energy $E_0$ into the calculations, confirm again the
anomaly of the intensity ratio $R$.  As the overall errors of the
keV-measurements were about 3\% while the actual statistical
errors were about 2\% \cite{Moreh}, one  concludes that the
revealed anomaly of ca. 20\% is highly significant.

\section{Theoretical remarks}

Originally it has been proposed \cite{PRL97} that the considered
effect is caused by short-lived and spatially restricted
entanglement. Since the typical neutron- and electron-proton
interaction time in NCS and ECS (i.e.~$\tau_{sc}$) lies within the
attosecond range, it is expected that decoherence may still not be
fully effective. Published theoretical models attribute this
effect to:

$(A)$ Modification of scattering due to "identity of particles" in
the scattering system. The influence on scattering of entanglement
of the spin and spatial degrees of freedom of identical particles
(i.e., quantum exchange correlations) has been stressed in
Refs.~\cite{Karlsson1}. In particular, NCS from pairs of protons
and deuterons has been calculated.

$(B)$ Contribution of electronic degrees of freedom to the
dynamics of a struck proton (deuteron)
 interacting with its environment.
  E.g., breakdown of
the Born-Oppenheimer (BO) approximation in the final state of the
NCS process and $(B.1)$  additional excitations  of the electronic
system \cite{Nikitas,GeorgeReiter}, and/or $(B.2)$  decoherence
accompanying short-lived spatial entanglement of a struck proton
with adjacent electrons and perhaps also nuclei \cite{Aris+Stig}.
In the models of this category, spin entanglement and/or quantum
exchange correlations between identical particles play no role.

To test certain contradictory predictions of these theories, we
recently measured by NCS
 $(a)$ the equimolar H$_2$--D$_2$
mixture and $(b)$ the mixed-isotope system HD (liquids, both at 20
K). The crucial result was that both systems reveal the same,
strong anomalous shortfall (about 30\%) of the ratio
$\sigma_H/\sigma_D$ of H and D cross-sections \cite{HD}. Since HD
exhibits no exchange correlations, this result
 demonstrates  for the first time  that these
correlations play no significant role in this effect, thus
refuting
corresponding theoretical models claiming its interpretation. This
conclusion is in line with a recently presented full calculation
of the scattering function $S(q,\omega)$ by Sugimoto et
al.~\cite{Sugimoto}, who found that indistinguishability of
particles cannot represent the physical origin of the observed
effect.

In contrast, our findings \cite{HD} are consistent with
alternative theoretical models of category $(B)$, in which
attosecond dynamics of electronic degrees of freedom  --- e.g. via
violation of the BO approximation --- is considered to participate
significantly to the dynamics of an elementary neutron-proton
(-deuteron) scattering process.  The scattering times of NCS and
ECS are similar to the characteristic time of "electron motion",
so that the Born-Oppenheimer approximation is not applicable here;
cf.~Ref.~[2a]. However, it should be stressed that a detailed
treatment of the BO failure and a quantitative estimate of its
possible contribution to the considered NCS-anomaly does not exist
yet, as convincingly discussed in Ref.~\cite{Daniele-NonBO}.

\section{Additional remarks}

The preceding derivations concerning the new keV experimental
results \cite{Moreh} have various far-reaching theoretical and
experimental consequences. Some of them are as follows.

First, it should be stressed that the intensity ratios determined
in this experiment can only provide information about a possible
\textit{difference} between the scattering behavior of H and D,
and not about that of H alone. In view of several theoretical
works \cite{Nikitas,GeorgeReiter,Aris+Stig}, it may be expected
that, due to the high energy transfers, both protons and deuterons
should violate the BO approximation, thus leading to the above
"anomaly" for both H and D. Thus, the measurement of the
scattering intensity ratio $R_{exp}$ of H$_2$O relative to that of
D$_2$O is not appropriate for the search of the anomalous
scattering effect under consideration. Obviously, an amended keV
experiment would be to measure the scattering intensity ratio of
H$_2$O (D$_2$O, and their mixtures) relative to that of a heavy
metal sample, e.g. Pb. This experiment was recently proposed by
the author.

Second, already earlier NCS from D-containing materials have shown
that also D exhibits a small anomalous shortfall of scattering
intensity. E.g. in [2c] was reported that this shortfall in
NbD$_{0.8}$ was about $10$\%. Furthermore, several NCS
measurements on pure D$_2$O showed up a shortfall of the cross
sections ratio $\sigma_D/\sigma_O$ of about 10-15\% (unpublished
data).
 Thus, the aforementioned 20\%-anomaly of
the measured ratio $R_{exp}$ in the keV experiment indicates that
the integrated scattering intensity of pure H$_2$O may exhibit an
anomalous shortfall larger than 20\%. In view of these
considerations, the aforementioned proposed experiment may be
expected to yield an anomalous intensity ratio of H$_2$O to Pb of
the order of 25\% and more.

Summarizing,
 we conclude that the considered scattering effect is
 present at both 5--100 eV \cite{PRL97} and 24--150 keV
 \cite{Moreh} ranges of incident energies.
 Obviously, the novel experiment \cite{Moreh} and its correct
analysis established above open up new perspectives for neutron
research on the above attosecond effect, and thus they may
have far reaching consequences for current and future experimental
and theoretical investigations.\\

 \section*{Acknowledgment}
I acknowledge
  partial support by the EU (the
QUACS RTN) and by a grant from the Royal Swedish Academy of
Sciences.

\end{document}